# Robust magnetoresistance in TaAs$_2$ under pressure up to about 37 GPa

Hongyuan Wang[1,2,3], Cuiying Pei[1], Hao Su[1,2], Zhenhai Yu[1,*], Mingtao Li[4], Wei Xia[1,2,3], Xiaolei Liu[1,2], Qifeng Liang[5*], Jinggeng Zhao[6], Chunyin Zhou[7], Na Yu[1,8], Xia Wang[1,8], Zhiqiang Zou[1,8], Lin Wang[4], Yanpeng Qi[1,*] and Yanfeng Guo[1,*]

[1] School of Physical Science and Technology, ShanghaiTech University, Shanghai 201210, China

[2] University of Chinese Academy of Sciences, Beijing 100049, China

[3] Shanghai Institute of Optics and Fine Mechanics, Chinese Academy of Sciences, Shanghai 201800, China

[4] Center for High Pressure Science and Technology Advanced Research, Shanghai, 201203, China

[5] Department of Physics, University of Shaoxing, Shaoxing 312000, China

[6] Department of Physics, Harbin Institute of Technology, Harbin 150080, China

[7] Shanghai Synchrotron Radiation Facility, Shanghai Advanced Research Institute, Chinese Academy of Sciences, Shanghai 201204, China

[8] Analytical Instrumentation Center, School of Physical Science and Technology, ShanghaiTech University, Shanghai 201210, China

*Corresponding authors:
yuzhh@shanghaitech.edu.cn,
qfliang@usx.edu.cn,
qiyp@shanghaitech.edu.cn,
guoyf@shanghaitech.edu.cn.

**Abstract**

  The extremely large magnetoresistance (XMR) in nonmagnetic semimetals has inspired growing interest owing to both intriguing physics and potential applications. We report results of synchrotron X-ray diffraction (SXRD) and electrical transport measurements on $TaAs_2$ under pressure up to ~ 37 GPa, which revealed an anisotropic compression of the unit cell, formation of unusual As-As bonds above 9.5 GPa, and enhancement of metallicity. Interestingly, the MR of $TaAs_2$ under pressure changed gently, which at 1.7 GPa is 96.6% and at 36.6 GPa is still 36.7%. The almost robust MR under pressure could be related to the nearly stable electronic structure unveiled by the *ab initio* calculations. The discovery would expand the potential use of XMR even under high pressure.

The tunable electrical resistance ($R$) of some materials by external magnetic field ($B$) is capable of producing the MR effect, where MR is defined as MR = [$\rho(B) - \rho(0)$]/$\rho(0)$ × 100%. The effect is usually very weak in conventional nonmagnetic semimetals because the compensated electron and hole carrier densities ($n_e/n_h \sim 1$) are rather low. Surprisingly, extremely large and nonsaturating MR with the magnitude even up to ~ $10^4$% - $10^6$% at moderate magnetic fields was discovered in some nonmagnetic semimetals such as WTe$_2$ [1], Cd$_3$As$_2$ [2], lanthanum monopnictides [3-5], PtSn$_4$ [6], MPn (M = Ta, and Nb, Pn = P and As) [6-10], and MPn$'_2$ (Pn$'$ = As and Sb) [11-15], etc. To interpret the XMR in these nonmagnetic semimetals, mainly two mechanisms have been proposed. One describes a quantum effect near the linearly dispersed low-energy bulk electrons in topologically protected band structure [16, 17]. The lift of the topological protection by external $B$ could give rise to linear field-dependent XMR, such as in the Dirac semimetal Cd$_3$As$_2$ and Na$_3$Bi [18, 19], and the Weyl semimetals MPn [6-10], etc. The other scenario relied on the isotropic semi-classical model with perfect electron-hole compensation is predicted to responsible for the large positive quadratic field-dependent MR (MR $\propto B^2$), such as in graphite, bismuth, MPn'$_2$ (Pn' = As and Sb), YbSb [4], LaSb [5], etc. It was supported by the angle-resolved photoemission spectroscopy (ARPES) measurements. However, origin of the XMR in some materials such as the type-II Weyl semimetal WTe$_2$ and the nodal-line semimetal CaTX (T = Ag, Cd; X= As, Ge) [20], has been still under debate between the two mechanisms. They both are even argued to be responsible for the XMR simultaneously [1].

The compensated semimetal TaAs$_2$ possess positive XMR reaching ~ $10^5$ - $10^6$ % at $B$ = 9 T and $T$ = 2 K [11]. Recent density functional calculations on it unveiled a new topological semimetal nature with the Z$_2$ invariant (0; 111) [15]. More interestingly, very large negative MR (~ −98%) that could be well fitted by the Adler-Bell-Jackiw (ABJ) chiral anomaly was also detected [15], likely challenging the widely accepted carrier compensation mechanism. On the other side, recent studies on NbAs$_2$ under pressure detected superconductivity with the $T_c$ of 2.63 K [21]. It naturally reminds us of the examination of possible superconductivity in pressured TaAs$_2$. Furthermore, to expand the practical use of the XMR of TaAs$_2$, the pressure stability definitely deserves an examination.

Crystals of TaAs$_2$ were grown by a chemical vapor transport method [11]. The crystallographic phase and quality were examined on a Bruker D8 VENTURE single crystal diffractometer using Mo $K_{\alpha 1}$ radiation ($\lambda$ = 0.7093 Å) at room temperature. Electrical transport measurements under HP were carried out using a BeCu Diamond Anvil Cell (DAC) with four-probe method in a 9 T DynaCool physical property measurement system. The HP synchrotron XRD experiments were performed at room temperature by using a symmetric DAC and T301 stainless steel gasket. The 120 μm diameter sample chamber was filled with a mixture of sample powder, a ruby chip, and silicone oil as the pressure-transmitting medium. Angle dispersive XRD (AD-XRD) experiments for TaAs$_2$ were performed at BL15U1 beamline (wavelength: 0.6199 Å) of the Shanghai Synchrotron Radiation Facility (SSRF). The AD-XRD experiments were carried out at room temperature. The pressure determination in our experiments was according to the fluorescence shift of ruby [22]. Fit2D software package was used to process the data [23]. The XRD patterns were analyzed with Rietveld refinement using the GSAS program package [24] with a user interface EXPGUI [25]. The first-principle calculations were performed by the Vienna *ab initio* simulation package (VASP) [26] and the projected augmented-wave (PAW) potential was adopted [27, 28]. The exchange-correlation functional introduced by Perdew, Burke, and Ernzerhof (PBE) within generalized gradient approximation (GGA) was applied in the calculations [29]. The energy cutoff for the plane-wave basis was set as 520 eV and the forces were relaxed less than 0.01 eV/Å. The positions of atoms were allowed to relax while the lattice constants of the unit cells were fixed to the experimental values.

The single crystal XRD analysis confirmed the monolithic NbAs$_2$-type structure (space group $C2/m$, no. 12) for TaAs$_2$ with $a$ = 9.3385 Å, $b$ = 3.3851 Å, $c$ = 7.7568 Å and $\beta$ = 119.70°( Figure S1 of the supporting information). Figure 1(a) presents selected AD-XRD patterns of TaAs$_2$ under various pressures. The Bragg peaks exhibit minute shift toward higher angles caused by the lattice contraction upon increasing pressure. The XRD measurements with the pressure up to 37.6 GPa did not detected any new diffraction peaks arising from other phases or impurities. The Rietveld refinement therefore adopted the $C2/m$ structure as initial model and the result of the data at 12.9 GPa is shown in Figure 1(b), confirming that the model is correct.

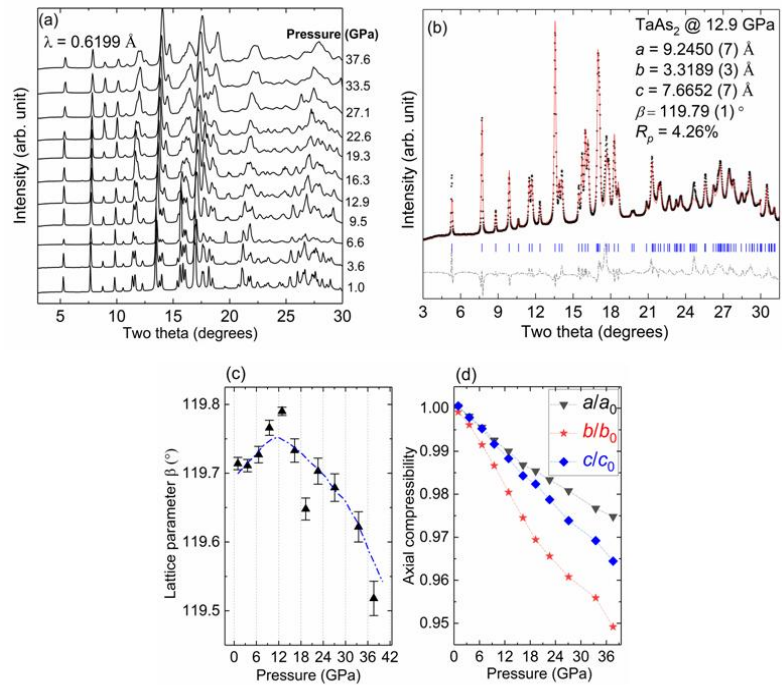

Figure 1. (a) The selected angle dispersive XRD patterns of TaAs$_2$ under various pressures up to 37.6 GPa at room temperature. (b) The Rietveld refinement results of TaAs$_2$ structure at 12.9 GPa. The vertical bars represent the calculated positions of the diffraction peaks. The difference between the observed (scatters) and the fitted patterns (line) is shown at the bottom of the diffraction peaks. (c) The pressure dependence of lattice parameters $\beta$ of TaAs$_2$. (d) The pressure dependence of the relative axial compressibility for TaAs$_2$.

The refined lattice parameters *a*, *b*, *c* and unit-cell volume (*V*) exhibit a monoclinic reduction upon pressure increasing (Figure S2 of supporting information). Seen in Figure 1(c), *β* exhibits an initial increase followed by a decrease with increased pressure. The pressure dependence of relative axial compressibility is depicted in Figure 1(d), revealing an anisotropic relative axial compressibility along different axes and the *b*-axis is more compressible.

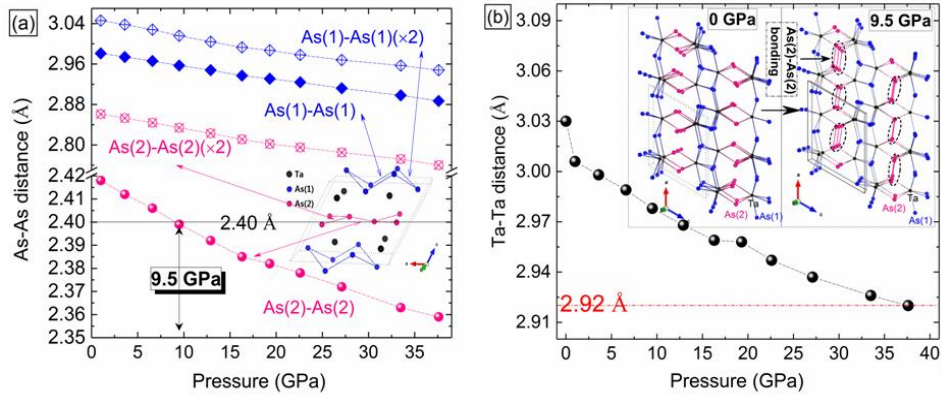

Figure 2. (a) Pressure dependence of selected interatomic distances of As-As in TaAs$_2$: the dashed lines with scatters are guides for the eye. Inset shows the nonequivalent As-As bondings in unit cell. The threshold value (2.4 Å) for formation of As-As bonding is indicated by the straight horizontal line at the bottom of figure. (b) Ta-Ta distance as a function of pressure. Inset shows the schematic formation of As(2)-As(2) bonding under high pressure.

The formation of unusual As-As bonding states in TaAs$_2$ under HP was observed. The shortest As-As distance at ambient condition in TaAs$_2$ is 2.42 Å, implying that there is no As-As bonding because the covalent radius of As is 1.2 Å [30]. Formation of chemical bonding in metalloid anions such as P, As and Sb was previously theoretically investigated in transition metal pnictides [31]. The experimental observation of a formation of As-As interlayer bonding was in the collapsed tetragonal NaFe$_2$As$_2$ under pressure [32]. The As-As distance for As(1) and As(2) atoms in pressured TaAs$_2$ is presented in Figure 2(a), showing that all these As-As distances decrease upon increasing pressure. The As(1)-As(1) distance at AP is ~ 3.0 Å and decreases gently with increasing the pressure, eventually reaches a minimal value of 2.887 Å at 37.6 GPa, indicating that the As(1)-As(1) bonding could not be formed. The arrangement of the As(2) atoms produces alternately longer (~ 2.87 Å) and shorter (~ 2.42 Å) As(2)-As(2) distance at AP. The short As(2)-As(2) distance in Figure 2(a) exhibits clear decreases with increasing pressure and becomes smaller than 2.40 Å at ~ 9.5 GPa, suggesting that the short As(2)-As(2) interactions (< 2.4 Å) are actually chemical bonds as shown by the inset in Figure 2(b). The bonding state between metalloid As atoms is one of the intriguing features and it tends to enhance the metallicity of TaAs$_2$. Furthermore, the Ta-Ta distance decreases upon the pressure increasing and approaches 2.92 Å at 37.6 GPa, as shown in Figure 2(b), which indicates that the formation of Ta-Ta bonding is at much higher pressure than that of As(2)-As(2).

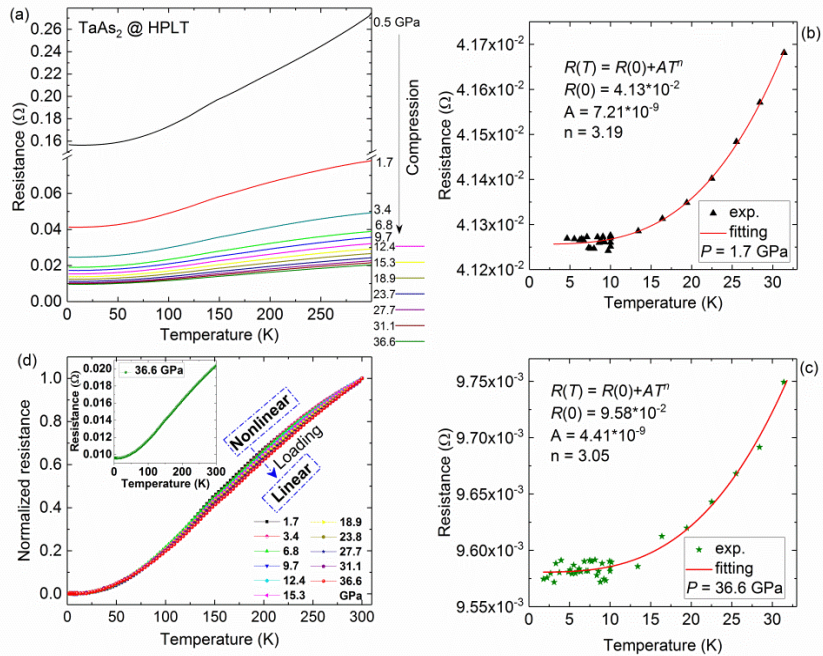

Figure 3. (a) Pressure dependence of R(T) for TaAs$_2$. (b)-(c) The fitting results of R(T) (2 K ~ 30 K) at (b) 1.7 GPa and (c) 36.6 GPa, respectively. (d) Pressure dependent normalized R(T) at selected pressures. Within 30 K ~ 300K, R(T) shows nonlinear to linear transformation indicated by the arrow with increased

pressure. Inset shows the R(T) at 36.6 GPa.

The pressure dependent $R(T)$ of TaAs$_2$ is presented in Figure 3(a), displaying monoclinic decrease in magnitude with increasing the pressure. The low temperature upturn of $R(T)$ at AP is easily completely suppressed with a small pressure. Unfortunately, unlike NbAs$_2$, no superconductivity in TaAs$_2$ was traced within the measured pressure range. The lower temperature $R(T)$ was fitted by the Bloch-Gruneissen equation expressed as $R(T) = R_0 + A \cdot T^n$ where $R_0$ is the residual resistivity at zero temperature, $A$ and $n$ are fitting parameters. The results of representative fitting to the data at 1.7 GPa and 36.6 GPa were presented in Figures 3(b)-3(c). A power law behavior is clearly visible with the exponent $n$ decreasing from 3.4 at 1.7 GPa to 3.0 at 36.6 GPa. The two values are close to that expected for *s-d* electron scattering, $n$ = 3 [33]. The results imply that inter-band scattering within Ta orbitals or between Ta and As orbitals plays crucial role in influencing the transport properties. At low pressures, we note that $R(T)$ of TaAs$_2$ show positive curvatures at high temperature region, seen in Figure 3(d), which is proposed to originate from the resistivity saturation term of the sample with electron mean free path $l$ being comparable with the lattice parameters [34]. With further increasing the pressure, $l$ may eventually exceed the progressively compressed lattice parameters, thus leading to the visible transition of $R(T)$ from the nonlinear to almost linear temperature dependent behavior as is indicated by the marked arrow.

The MR of TaAs$_2$ at 2 K and 9 T is ~ $2.1 \times 10^5$%, which shows clear suppression by the application of pressure. The value is 96.6% at 2 K, 9 T and 1.7 GPa but then surprisingly changes gently with continuously increased pressure, which is 90.9%, 81.8%, 77.4% and 36.7% at 3.4 GPa, 6.8 GPa, 9.7 GPa, and 36.6 GPa, respectively, seen in Figure 4(a). Li *et al*. reported that when a 14.6 GPa pressure is applied, MR of NbAs$_2$ at 2 K and 8 T is effectively suppressed to 8% of that at AP and is almost inhibited with increasing the pressure to 14.6 GPa [21]. Our result implies a robust MR of TaAs$_2$ under pressure. The power-law relation, $MR = (\bar{\mu}B)^\alpha$ where the $\bar{\mu}$ is the averaged mobility and $\alpha$ is a fitting coefficient, was used to fit the experimental data as shown in Figure 4(b), revealing that the averaged carrier mobility clearly decreases with increasing the pressure, which should play an import role in reducing the MR.

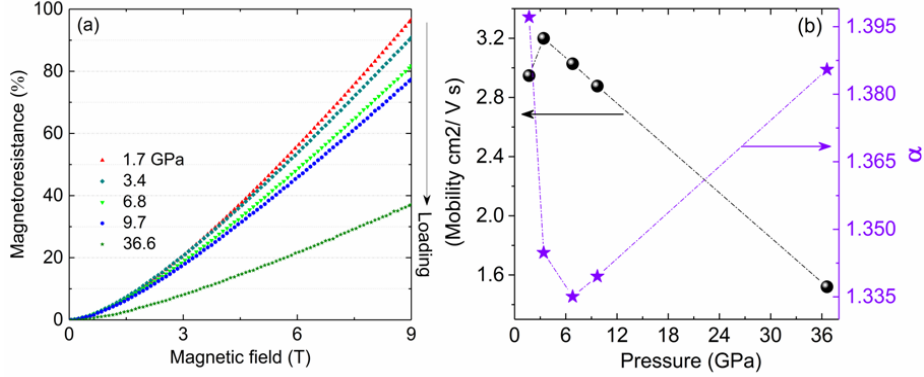

Figure 4. Magnetoresistance of TaAs$_2$ at 2 K and 9 T is slowly suppressed with increasing the pressure up to 36.6 GPa. (b) the pressure dependence of averaged mobility.

The calculated band structures are presented in Figure 5. Under ambient condition and without the spin-orbital coupling (SOC), the conduction and valence bands of TaAs$_2$ cross along three high-symmetry paths $I_1$–$Z$, $I$–$L$ and $X_1$–$Y$ (see in Figure 5(a)), while at other paths the band dispersion is relatively small which can not induce a band-inversion. One can easily confirm that these band-crossing paths are parallel to the double-chains. Because of the chain-like structure of TaAs$_2$, the electrons move more freely along the chains than between the chains, which explains the large-dispersions and band-crossings at $I_1$–$Z$, $I$–$L$ and $X_1$–$Y$. When SOC is included, the band crossings at $I_1$–$Z$, $I$–$L$ and $X_1$–$Y$ are lifted and energy gaps open at these crossing points. Previous studies have classified the TaAs$_2$ as a type-II Dirac semimetal without SOC and a weak topological insulator with SOC by computing the $Z_2$ indices at the time-reversal-invariant-momenta of the Brriloiun zone (BZ) [8,9].

Next, we show that electronic structure of TaAs$_2$ is only slightly altered by the HP, and the conduction and valence bands are kept in being inverted. In Figures 5 (b) and 5(c), we show the band structures of TaAs$_2$ at the pressure of 9.5 and 37.6 GPa, respectively. Due to the shrinking of lattice constants, one sees the band dispersion of electrons at $I_1$–$Z$, $I$–$L$ and $X_1$–$Y$ increases as compared to the band structure of TaAs$_2$ at ambient conditions in Figure 5(a), which even further enhances the band inversion at these paths. Therefore, we can conclude that the electronic structure of TaAs$_2$ is highly stable against the external pressure and the material remains a weak topological insulator.

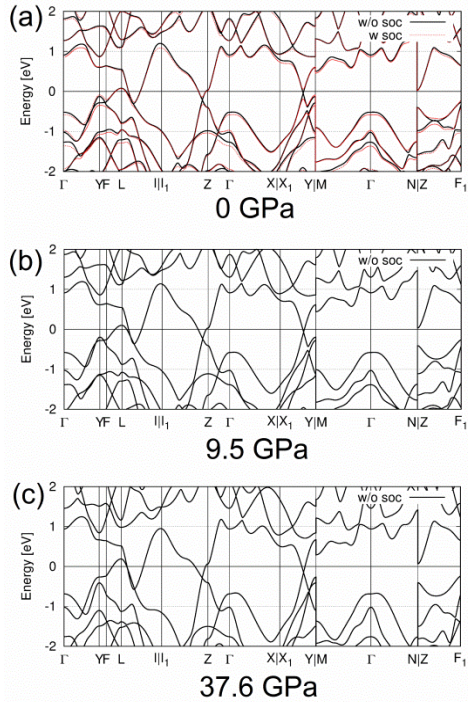

Figure 5. Band structures of TaAs$_2$ at different pressures.

As a summary, we report observation of the formation of unusual As-As bonding and the nearly robust XMR in TaAs$_2$ under high pressure. The formation of As-As bonding highlights the role of high pressure in creating unusual bonding states that would influence the physical properties or give rise to exotic behaviors. Considering the potential applications of the NbAs$_2$-family of materials, the robust XMR makes TaAs$_2$ more promising for use even under HP. The results would also bring valuable clues for understanding the electrical transport properties and intriguing band topology of TaAs$_2$.

The authors acknowledge the support by the Natural Science Foundation of Shanghai (Grant No. 17ZR1443300), the Shanghai Pujiang Program (Grant No. 17PJ1406200), the National Natural Science Foundation of China (Grant No. 11874264) and the Natural Science Foundation of Heilongjiang Province (Grant No. A2017004). We thank Lili Zhang and Shuai Yan (from BL15U1 beamline, Shanghai Synchrotron Radiation Facility in China.) for help in high-pressure synchrotron X-ray diffraction (SXRD) measurements.

# *Supporting information*

## Robust magnetoresistance in TaAs$_2$ under pressure up to about 37 GPa

Hongyuan Wang[1,2,3], Cuiying Pei[1], Hao Su[1,2], Zhenhai Yu[1,*], Mingtao Li[4], Wei Xia[1,2,3], Xiaolei Liu[1,2], Qifeng Liang[5*], Jinggeng Zhao[6], Chunyin Zhou[7], Na Yu[1,8], Xia Wang[1,8], Zhiqiang Zou[1,8], Lin Wang[4], Yanpeng Qi[1,*] and Yanfeng Guo[1,*]

[1]School of Physical Science and Technology, ShanghaiTech University, Shanghai 201210, China

[2]University of Chinese Academy of Sciences, Beijing 100049, China

[3]Shanghai Institute of Optics and Fine Mechanics, Chinese Academy of Sciences, Shanghai 201800, China

[4]Center for High Pressure Science and Technology Advanced Research, Shanghai, 201203, China

[5]Department of Physics, University of Shaoxing, Shaoxing 312000, China

[6]Department of Physics, Harbin Institute of Technology, Harbin 150080, China

[7]Shanghai Synchrotron Radiation Facility, Shanghai Advanced Research Institute, Chinese Academy of Sciences, Shanghai 201204, China

[8]Analytical Instrumentation Center, School of Physical Science and Technology, ShanghaiTech University, Shanghai 201210, China

*Corresponding authors:

yuzhh@shanghaitech.edu.cn,

qfliang@usx.edu.cn,

qiyp@shanghaitech.edu.cn,

guoyf@shanghaitech.edu.cn.

## The crystal sructure of $TaAs_2$ under ambient conditions

There are two chemical sites for As atoms in each unit cell, labeled as As(1) and As(2), and Ta atoms are located inside a trigonal prism formed by the As atoms. The prisms are connected in pairs by sharing one square face. The Ta-As prisms are vertex-connected along the *a*-axis and edge-shared along the *c*-axis. The neighboring Ta-As layers along the *b*-axis are connected in face-shared configuration. The prisms are stacked along the crystallographic *b* direction through their trigonal faces, implying possible anisotropic compressibility along different crystallographic directions. At ambient conditions, each Ta atom in $TaAs_2$ is surrounded by six As atoms at the corners of a triangular prism and by two other As atoms and one Ta outside the rectangular faces of the prism, shown by the inset of Figure S1(a). The As(1) atoms are surrounded by five near-neighboring Ta in a distorted square pyramidal arrangement and by one close-neighboring As (1) located across the base of the pyramid. The As(2) atoms are coordinated to three As(2) atoms and three Ta atoms.

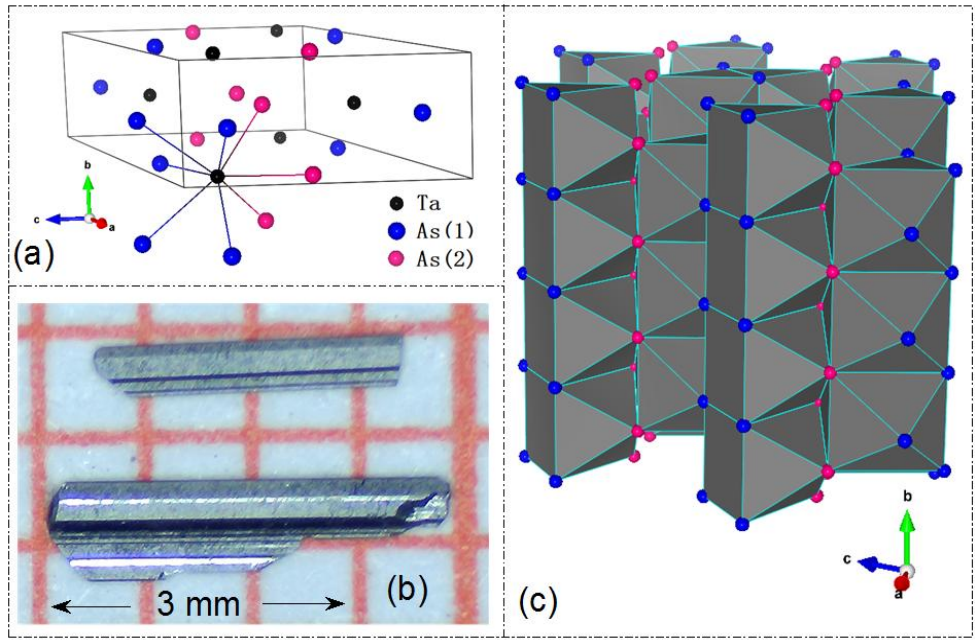

Figure S1. (a) depicts the schematic crystal structure of $TaAs_2$. (b) shows optical image of the typical $TaAs_2$ single crystal used in this work. (c) illustrates the stacking sequence of Ta-As prisms along the *b*-axis.

## The pressure dependence of lattice parameters for TaAs$_2$

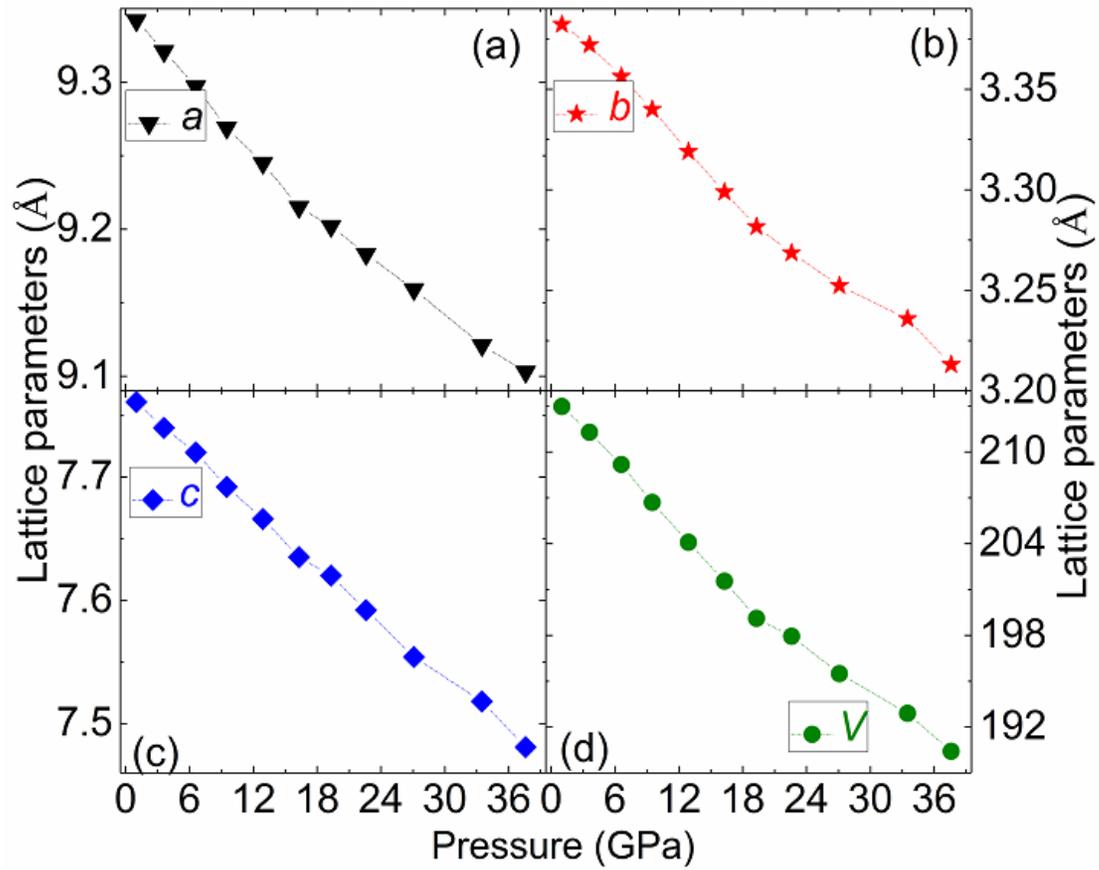

Figure S2. The pressure dependence of lattice parameters *a* (a), *b* (b), *c* (c) and *V* (d) of TaAs$_2$.